\documentclass[article,twocolumn]{revtex4-1}

\usepackage{amsmath,xcolor}
\usepackage{amsfonts}
\usepackage{amssymb}
\usepackage{graphicx}
\usepackage{srcltx}
\usepackage{mathcomp} %allows "per mille" symbol through the command \tcperthousand
\usepackage{ulem} %allows to strike out the text with \sout

\newcommand{\beq}{\begin{equation}}
\newcommand{\eeq}{\end{equation}}
\newcommand{\ket} [1] {|#1\rangle}
\newcommand{\bra} [1] {\langle#1|}
\newcommand{\braket}[2]{\langle #1 | #2 \rangle}

\begin{document}

\title{Reply to ``Comment on 'Strong Measurements Give a
Better Direct Measurement of the Quantum
Wave Function' "}

%%%%%%%%%%%%%%%%%%%%%%%%%%%%%%%%%%%%%%%%%%%%%%%%%%%%%%%%%%%%%%%%%%%%%%%%%%%
%
\author{Giuseppe Vallone$^{1}$}
%\email{vallone@dei.unipd.it}
%
\author{Daniele Dequal$^{2}$}
%
%
%%%%%%%%%%%%%%%%%%%%%%%%%%%%%%%%%%%%%%%%%%%%%%%%%%%%%%%%%%%%%%%%%%%%%%%%%%%
% 1
\affiliation{$^1$Department of Information Engineering, University of Padova, I-35131 Padova, Italy}
% 2
\affiliation{$^2$Italian Space Agency, ASI, Rome, Italy}
% 3
%
%%%%%%%%%%%%%%%%%%%%%%%%%%%%%%%%%%%%%%%%%%%%%%%%%%%%%%%%%%%%%%%%%%%%%%%%%%%
%
\begin{abstract}
\end{abstract}
%
%%%%%%%%%%%%%%%%%%%%%%%%%%%%%%%%%%%%%%%%%%%%%%%%%%%%%%%%%%%%%%%%%%%%

\maketitle

%%%%%%%%%%%%%%%%%%%%%%%%%%%%%%%%%%%%%%%%%%%%%%%%%%%%%%%%%%%%%%%%%%%

\noindent In Ref.~\cite{vallone16prl}, we 
generalized the method introduced in Ref.~\cite{lundeen11nat} to obtain a direct measurement of a quantum wave-function.
The method of Ref.~\cite{lundeen11nat} is based on  weak measurements~\cite{aharonov88prl} to determine the wave-function.
In Ref.~\cite{vallone16prl} we showed that weak measurements are not necessary and
indeed a better direct measurement (in term of precision and accuracy) of the quantum wave-function can be obtained by using strong measurements.

Recently, our work received a comment by V. R. Pande, S. Kanjilal and D. Das (PKD) \cite{pande17qph} about a flaw our method.
The argument of PKD can be summarized in one of their statements:
\begin{quotation}
``we show that certain entities in Eq. (S6) of their supplemental materials on which the central claim of their result is based have no operational existence and are therefore not experimentally measurable.''
\end{quotation}
In this reply, we show that their claim is not correct and that the quantities appearing in our method can be indeed measured.

As correctly stated in \cite{pande17qph}, the probabilities needed
for our method (namely $P^{(x)}_+$, $P^{(x)}_-$, $P^{(x)}_1$, $P^{(x)}_R$ and $P^{(x)}_R$, 
defined in the supplementary material of \cite{vallone16prl}) 
are calculated by using the unnormalized pointer wave-function $\ket{\varphi}_P$ (Eq. (S2)) such as
\beq
\label{Pj} 
P^{(x)}_j=|\braket{j}{\varphi}|^2\,, 
\eeq
with $j=+,-,1,R,L$ and the pointer states defined as follows
\beq
\begin{aligned}
\ket{+}_P&=\frac1{\sqrt2}(\ket0+\ket1)\,,\quad&
\ket{-}_P&=\frac1{\sqrt2}(\ket0-\ket1)
\\
\ket{L}_P&=\frac1{\sqrt2}(\ket0+i\ket1)\,,\quad&
\ket{R}_P&=\frac1{\sqrt2}(\ket0-i\ket1)\,.
\end{aligned}
\eeq
In the above equations, $\{\ket0,\ket1\}$ is a basis for the pointer Hilbert space.
The unnormalized pointer state is defined as
\beq
\ket{\varphi}_P=\ _X\bra{p_0}U_x(\theta)\ket{\psi}_X\otimes\ket0_P\,,
\eeq
with $\ket0_P$ the initial state of the pointer, $\ket{\psi}_X$  the unknown state whose wave-function will be determined,
$U_x(\theta)$  the interaction unitary and $\ket{p_0}_X=\frac1{\sqrt d}\sum_x\ket x$ the zero transverse momentum state.
Since $\ket\varphi$ is not normalized, clearly $P^{(x)}_++P^{(x)}_-=\braket{\varphi}{\varphi}\neq 1$.

The mistake of PKD is to consider that the quantities $P^{(x)}_j$ should be measured on the post-selected quantum ensemble
and  that they correspond to conditional probabilities (as in the original proposal \cite{lundeen11nat}).

Indeed, as shown below, the quantities $P^{(x)}_j$ are not {\it conditional probabilities} but {\it joint probabilities}.
By defining $\ket{\Psi'}=U_x(\theta)\ket{\psi}_X\otimes\ket0_P$ as the joint system-pointer state after the interaction,
it is easy to show that the probabilities in eq. \eqref{Pj} can be written as
\beq 
\label{Pj2}
P^{(x)}_j=|\bra{j}\otimes\braket{p_0}{\Psi'}|^2\equiv {\rm Prob}_{\Psi'}(j,p_0)\,.
\eeq
The above equation shows that the quantities $P^{(x)}_j$ represent the {\it joint probabilities} of measuring
the system in the state $\ket{p_0}$ and the pointer in the state $\ket{j}$ after the interaction.
Such probabilities are well defined, can be measured and thus have ``operational existence''.

On the other hand, probabilities measured on the post-selected quantum ensemble
(after the post-selection on the state $\ket{p_0}$)
are {\it conditional probabilities}, and indeed they can be obtained by Bayes' rule as
\beq
 {\rm Prob}_{{\Psi'}}(j|p_0)= \frac{{\rm Prob}_{{\Psi'}}(j,p_0)}{P(p_0)}=\frac{P^{(x)}_j}{\braket{\varphi}{\varphi}}\,.
 \eeq
We note that the normalization factor $\braket{\varphi}{\varphi}$ corresponds to the probability of post-selection, since
\beq
\braket{\varphi}{\varphi}=\bra{p_0}\text{Tr}_P\big[\ket{\Psi'}\bra{\Psi'}\big]\ket{p_0}=P_{\rm post-selection}\,.
\eeq

The mistake of PKD is claiming that only the conditional probabilities $ {\rm Prob}_{\ket{\Psi'}}(j|p_0)$ can be measured
while the probabilities that are ``calculated from an unnormalized pointer state [...] cannot be measured experimentally''.
As we have shown in eq. \eqref{Pj2}, the probabilities $P^{(x)}_j$ obtained from the unnormalized state $\ket\varphi_P$
are {\it joint probabilities} that can be easily measured in experiments. By using joint probabilities of eq. \eqref{Pj2} it is possible to 
show that the 
wavefunction $\psi_x=\braket{x}{\psi}$ can
be expressed as
\beq
\begin{split}
\psi_x=\frac{d}{2\widetilde \psi\sin\theta}
\left[P^{(x)}_+-P^{(x)}_-+2P^{(x)}_1\tan\frac\theta2+\right.
\\
\left.+i(P^{(x)}_L-P^{(x)}_R)\right]
\end{split}
\eeq
with $\widetilde \psi=\sum_x\psi_x$.
In the above equation, the normalization factor  
$\frac{d}{2\widetilde \psi\sin\theta}$ does not depend on $x$, and therefore it can be 
determined at the end of the procedure 
by imposing the overall normalization of the wave-function. 

It is worth noticing that any experiment designed to measure the conditional probabilities is also able to measure
joint conditional probabilities.

%

%\bibliography{library}% Produces the bibliography via BibTeX.

\begin{thebibliography}{4}%
\makeatletter
\providecommand \@ifxundefined [1]{%
 \@ifx{#1\undefined}
}%
\providecommand \@ifnum [1]{%
 \ifnum #1\expandafter \@firstoftwo
 \else \expandafter \@secondoftwo
 \fi
}%
\providecommand \@ifx [1]{%
 \ifx #1\expandafter \@firstoftwo
 \else \expandafter \@secondoftwo
 \fi
}%
\providecommand \natexlab [1]{#1}%
\providecommand \enquote  [1]{``#1''}%
\providecommand \bibnamefont  [1]{#1}%
\providecommand \bibfnamefont [1]{#1}%
\providecommand \citenamefont [1]{#1}%
\providecommand \href@noop [0]{\@secondoftwo}%
\providecommand \href [0]{\begingroup \@sanitize@url \@href}%
\providecommand \@href[1]{\@@startlink{#1}\@@href}%
\providecommand \@@href[1]{\endgroup#1\@@endlink}%
\providecommand \@sanitize@url [0]{\catcode `\\12\catcode `\$12\catcode
  `\&12\catcode `\#12\catcode `\^12\catcode `\_12\catcode `\%12\relax}%
\providecommand \@@startlink[1]{}%
\providecommand \@@endlink[0]{}%
\providecommand \url  [0]{\begingroup\@sanitize@url \@url }%
\providecommand \@url [1]{\endgroup\@href {#1}{\urlprefix }}%
\providecommand \urlprefix  [0]{URL }%
\providecommand \Eprint [0]{\href }%
\providecommand \doibase [0]{http://dx.doi.org/}%
\providecommand \selectlanguage [0]{\@gobble}%
\providecommand \bibinfo  [0]{\@secondoftwo}%
\providecommand \bibfield  [0]{\@secondoftwo}%
\providecommand \translation [1]{[#1]}%
\providecommand \BibitemOpen [0]{}%
\providecommand \bibitemStop [0]{}%
\providecommand \bibitemNoStop [0]{.\EOS\space}%
\providecommand \EOS [0]{\spacefactor3000\relax}%
\providecommand \BibitemShut  [1]{\csname bibitem#1\endcsname}%
\let\auto@bib@innerbib\@empty
%</preamble>
\bibitem [{\citenamefont {Vallone}\ and\ \citenamefont
  {Dequal}(2016)}]{vallone16prl}%
  \BibitemOpen
  \bibfield  {author} {\bibinfo {author} {\bibfnamefont {G.}~\bibnamefont
  {Vallone}}\ and\ \bibinfo {author} {\bibfnamefont {D.}~\bibnamefont
  {Dequal}},\ }\href {\doibase 10.1103/PhysRevLett.116.040502} {\bibfield
  {journal} {\bibinfo  {journal} {Physical Review Letters}\ }\textbf {\bibinfo
  {volume} {116}},\ \bibinfo {pages} {040502} (\bibinfo {year}
  {2016})}\BibitemShut {NoStop}%
\bibitem [{\citenamefont {Lundeen}\ \emph {et~al.}(2011)\citenamefont
  {Lundeen}, \citenamefont {Sutherland}, \citenamefont {Patel}, \citenamefont
  {Stewart},\ and\ \citenamefont {Bamber}}]{lundeen11nat}%
  \BibitemOpen
  \bibfield  {author} {\bibinfo {author} {\bibfnamefont {J.~S.}\ \bibnamefont
  {Lundeen}}, \bibinfo {author} {\bibfnamefont {B.}~\bibnamefont {Sutherland}},
  \bibinfo {author} {\bibfnamefont {A.}~\bibnamefont {Patel}}, \bibinfo
  {author} {\bibfnamefont {C.}~\bibnamefont {Stewart}}, \ and\ \bibinfo
  {author} {\bibfnamefont {C.}~\bibnamefont {Bamber}},\ }\href {\doibase
  10.1038/nature10120} {\bibfield  {journal} {\bibinfo  {journal} {Nature}\
  }\textbf {\bibinfo {volume} {474}},\ \bibinfo {pages} {188} (\bibinfo {year}
  {2011})}\BibitemShut {NoStop}%
\bibitem [{\citenamefont {Aharonov}\ \emph {et~al.}(1988)\citenamefont
  {Aharonov}, \citenamefont {Albert},\ and\ \citenamefont
  {Vaidman}}]{aharonov88prl}%
  \BibitemOpen
  \bibfield  {author} {\bibinfo {author} {\bibfnamefont {Y.}~\bibnamefont
  {Aharonov}}, \bibinfo {author} {\bibfnamefont {D.~Z.}\ \bibnamefont
  {Albert}}, \ and\ \bibinfo {author} {\bibfnamefont {L.}~\bibnamefont
  {Vaidman}},\ }\href@noop {} {\bibfield  {journal} {\bibinfo  {journal} {Phys.
  Rev. Lett.}\ }\textbf {\bibinfo {volume} {60}},\ \bibinfo {pages} {1351}
  (\bibinfo {year} {1988})}\BibitemShut {NoStop}%
\bibitem [{\citenamefont {Pande}\ \emph {et~al.}(2017)\citenamefont {Pande},
  \citenamefont {Kanjilal},\ and\ \citenamefont {Das}}]{pande17qph}%
  \BibitemOpen
  \bibfield  {author} {\bibinfo {author} {\bibfnamefont {V.~R.}\ \bibnamefont
  {Pande}}, \bibinfo {author} {\bibfnamefont {S.}~\bibnamefont {Kanjilal}}, \
  and\ \bibinfo {author} {\bibfnamefont {D.}~\bibnamefont {Das}},\ }\href@noop
  {} {\  (\bibinfo {year} {2017})},\ \Eprint {http://arxiv.org/abs/1711.00764}
  {arXiv:1711.00764} \BibitemShut {NoStop}%
\end{thebibliography}

\end{document}